\documentclass[a4paper]{jpconf}
\usepackage{graphicx}
\begin{document}
\title{The Last Neutrino Mixing angle $\theta_{13}$}

\author{A.B. Balantekin}

\address{Physics Department, University of Wisconsin, Madison, WI  53706 USA}

\ead{baha@physics.wisc.edu}

\begin{abstract}
Among the still unmeasured neutrino properties, the third neutrino mixing angle, $\theta_{13}$, is likely to be the one we will next find out. In this contribution, first a brief summary of the limits and the preliminary measurements of this angle is given. Second a critical assessment of a widely used formula connecting two- and three-flavor evolution is provided. 
\end{abstract}

\section{Introduction}

Solar, accelerator, and reactor experiments performed during the last two decades firmly established 
that neutrinos mix. In general, for three flavors, the neutrino mixing matrix contains three mixing angles, one CP-violating phase and two Majorana phases. The Majorana phases are accessible only in double beta decay experiments. Two of the mixing angles are well measured. From solar and reactor experiments we know the value of one of them (usually denoted $\theta_{12}$). In the atmospheric neutrino experiments we measured the other one (usually denoted $\theta_{23}$). Contrary to the original (and naive) expectations, both angles are large. In fact one gets $\theta_{23} \sim \pi/4$.  To have the complete information on neutrino mixing, we still need to know three more pieces of information: i) the value of the remaining mixing angle, $\theta_{13}$, ii) the mass hierarchy; and iii) the value of the CP-violating phase. Of these three, the remaining mixing angle is likely to be the next piece to be measured. 

One popular parametrization of the neutrino mixing matrix is 
\begin{equation}
\label{1}
 {\bf T}_{23}{\bf T}_{13}{\bf T}_{12}  = 
\left(
\begin{array}{ccc}
 1 & 0  & 0  \\
  0 & C_{23}   & S_{23}  \\
 0 & -S_{23}  & C_{23}  
\end{array}
\right)
\left(
\begin{array}{ccc}
 C_{13} & 0  & S_{13} e^{-i\delta_{CP}}  \\
 0 & 1  & 0  \\
 - S_{13} e^{i \delta_{CP}} & 0  & C_{13}  
\end{array}
\right) 
\left(
\begin{array}{ccc}
 C_{12} & S_{12}  & 0  \\
 - S_{12} & C_{12}  & 0  \\
0  & 0  & 1  
\end{array}
\right)
\end{equation}
where $C_{ij} = \cos \theta_{ij}$, $S_{ij} = \sin \theta_{ij}$, and $\delta_{CP}$ is the CP-violating phase.  Note that although the CP-violating phase multiplies $S_{13}$ in Eq. (\ref{1}), there is nothing special about this choice; one could have placed $\delta$ elsewhere in the mixing matrix. In fact the combination 
\begin{equation}
\label{2}
P( \bar{\nu}_{\mu} \rightarrow \bar{\nu}_e) - P( \nu_{\mu} \rightarrow \nu_e), 
\end{equation}
which quantifies the CP-violation in neutrino oscillations, is proportional to the sines of {\it all} the mixing angles. However, since we know the other mixing angles are {\it non-zero}, observation of CP-violation in neutrino oscillations hinges on a non-zero value of $\theta_{13}$.  

The current limit on $\theta_{13}$ quoted by the Particle Data Group is $\sin^2(2 \theta_{13}) < 0.15$ at 90\% C.L. \cite{Nakamura:2010zzi}. 
There are strong hints for a non-zero value of $\theta_{13}$. Already several years ago, it was pointed out that the slight discrepancy between solar neutrino experiments and the KamLAND reactor neutrino experiment could be explained by a small, but non-zero value of $\theta_{13}$ \cite{Balantekin:2008zm}. 
Indeed global fits indicate a slight preference for a non-zero value of $\theta_{13}$  \cite{Fogli:2008jx}. 
(A recent brief summary of the $\theta_{13}$ phenomenology is given in \cite{Mezzetto:2010zi}). Recent low-threshold analysis of the Sudbury Neutrino Observatory yields $\theta_{13} = 7.2 ^{+2.0}_{-2.8}$ degrees, albeit with non-Gaussian errors \cite{Aharmim:2009gd}. A global analysis by the KamLAND  collaboration indicates $\sin^2 \theta_{13} = 0.009^{+0.013}_{-0.007}$ at the 79\% C.L. 
\cite{Gando:2010aa}. Currently three reactor neutrino experiments (Double Chooz \cite{Ardellier:2006mn}, Daya Bay \cite{Guo:2007ug}, and RENO \cite{:2010vy}) as well as the T2K accelerator experiment 
\cite{Abe:2011ks} are working on a measurement of this angle. So far the best hint for a non-zero value of $\theta_{13}$ came from the T2K experiment which announced a 2.5$\sigma$ result: $0.03 < \sin^2 2\theta_{13} < 0.28$ for the normal and  $0.04 < \sin^2 2\theta_{13} < 0.34$ for the inverted hierarchies, both for $\delta_{CP}=0$ \cite{:2011sj}. 

\section{Reduction of three-flavor evolution}

To extract $\theta_{13}$ from solar neutrino data many times the formula
\begin{equation}
\label{b1}
P_{3\times3}( \nu_e \rightarrow  \nu_e) = \cos^4{\theta_{13}} \> 
P_{2\times2}( \nu_e \rightarrow  \nu_e \>{\rm with}\> N_e
\cos^2{\theta_{13}})  + \sin^4{\theta_{13}} 
\end{equation}
is used. Since many fits to solar neutrino data that give a non-zero value of of $\theta_{13}$ use 
Eq. (\ref{b1}), here we wish to give an assessment of the accuracy of this formula. 
To do so, following Ref. \cite{Balantekin:1999dx}, we introduce the
combinations 
\begin{eqnarray}
\label{rot1}
\tilde{\Psi}_{\mu} &=& \cos{\theta_{23}} \Psi_{\mu} -
\sin{\theta_{23}} \Psi_{\tau}, \\
\tilde{\Psi}_{\tau} &=& \sin{\theta_{23}} \Psi_{\mu} +
\cos{\theta_{23}} \Psi_{\tau},
\end{eqnarray}
which leads to the MSW evolution equation
\begin{equation} 
\label{CProt}
i \frac{\partial}{\partial t} 
\left(
\begin{array}{c}
  \Psi_e \\
 \tilde{ \Psi}_{\mu} \\
  \tilde{\Psi}_{\tau} 
\end{array}
\right) 
= \tilde{\bf H} 
\left(
\begin{array}{c}
  \Psi_e \\
  \tilde{\Psi}_{\mu} \\
  \tilde{\Psi}_{\tau} 
\end{array}
\right) 
\end{equation}
where 
\begin{equation}
\label{htilde}
\tilde{\bf H} = 
{\bf T}_{13}{\bf T}_{12} 
\left(
\begin{array}{ccc}
E_1  & 0  & 0  \\
0  & E_2  & 0  \\
0  &  0 & E_3  
\end{array}
\right) {\bf T}^{\dagger}_{12}{\bf T}^{\dagger}_{13}  + 
\left(   
\begin{array}{ccc}
 V_{e \mu} & 0  & 0  \\
 0 & 0 & 0  \\
0  & 0   & 0  
\end{array}
\right) .
\end{equation}
In writing Eq. (\ref{htilde}), since it is proportional to the identity, the neutral current contribution is dropped and 
\begin{equation}
  \label{wolfen1}
V_{e\mu} (x) = \sqrt{2} G_F  N_e (x) .
\end{equation}

To write the evolution operator associated with Eq. (\ref{CProt}), 
\begin{equation}
\label{CProtev}
i \frac{\partial}{\partial t} \tilde{\bf U}
= \tilde{\bf H} \tilde{\bf U} ,
\end{equation}
following the procedure of Ref. \cite{Balantekin:2003dc} we introduce the evolution operator ${\bf U}$: 
\begin{equation}
\label{a1}
\tilde{\bf U} = {\bf T}_{13} {\bf U} {\bf T}^{\dagger}_{13} 
\end{equation}
which satisfies the equation
\begin{equation}
\label{a2}
i \frac{\partial}{\partial t} {\bf U}
= {\cal H} {\bf U}
\end{equation}
with 
\begin{equation}
\label{h0}
{\bf T}^{\dagger}_{13} \tilde{\bf H} {\bf T}_{13} = 
{\cal H} = \left(\matrix{
     \frac{1}{2} \tilde{V} - \Delta_{21} \cos 2 \theta_{12}&
     \frac{1}{2} \Delta_{21} \sin 2 \theta_{12} &  \frac{1}{2} V_{\mu e}
     \sin 2 \theta_{13} \cr
     \frac{1}{2} \Delta_{21} \sin 2 \theta_{12} & -
     \frac{1}{2}\tilde{V} + \Delta_{21} \cos 2 \theta_{12}   & 0 \cr
      \frac{1}{2} V_{\mu e} \sin 2 \theta_{13}  & 0 &
     \frac{1}{2}(\Delta_{31}+\Delta_{32}) + V_{\mu e} - \frac{3}{2} 
     \tilde{V} }\right),
\end{equation}
where we took the CP-violating phase to be zero and introduced the modified matter potential 
\begin{equation}
\label{modpotent}
\tilde{V} = V_{\mu e} \cos^2\theta_{13}
\end{equation}
as well as the quantity 
\begin{equation}
\label{Delta}
\Delta_{ij} = \frac{m_i^2 - m_j^2}{2E} = \frac{\delta m_{ij}^2}{2E}. 
\end{equation}
Writing the Hamiltonian in Eq. (\ref{h0}) as
\begin{equation}
\label{a3}
{\cal H} = {\cal H}_0 + {\cal H}_1
\end{equation}
with
\begin{equation}
\label{a4}
{\cal H}_0 = \left(\matrix{
     \frac{1}{2} \tilde{V} - \Delta_{21} \cos 2 \theta_{12}&
     \frac{1}{2} \Delta_{21} \sin 2 \theta_{12} &  0 \cr
     \frac{1}{2} \Delta_{21} \sin 2 \theta_{12} & -
     \frac{1}{2}\tilde{V} + \Delta_{21} \cos 2 \theta_{12}   & 0 \cr
      0 & 0 &
     \frac{1}{2}(\Delta_{31}+\Delta_{32}) + V_{\mu e} - \frac{3}{2} 
     \tilde{V} }\right),
\end{equation}
and 
\begin{equation}
\label{a5}
{\cal H}_1 =  \frac{1}{2}  \sin 2 \theta_{13} V_{\mu e}
\left(\matrix{
  0 & 0 & 1 \cr
    0 & 0  & 0 \cr
      1  & 0 & 0 }\right),
\end{equation}
we see that Eq. (\ref{a2}) can be solved perturbatively, assuming $\theta_{13}$ is small. (In seeking such solutions we will take all the evolution operators at the neutrino production (t=0) to be the identity operator).  The contributions which are lowest order in $\theta_{13}$ will come from from the quantity $\tilde{V}$ and from the transformation in Eq. (\ref{a1}). Eq.(\ref{a4}) implies that if $\theta_{13}$ were zero, 
$\tilde{\Psi}_{\tau}$ would decouple from the other flavors. Writing ${\bf U} = {\bf U}_0 {\bf U}_1$, the solution of 
\begin{equation}
\label{a6}
i \frac{\partial}{\partial t} {\bf U}_0
= {\cal H}_0 {\bf U}_0
\end{equation}
is
\begin{equation}
\label{a7}
{\bf U}_0 =  \left(
\begin{array}{cc}
 {\bf S} &   0  \\
 0  & \beta  
\end{array}
\right), 
\end{equation}
where ${\bf S}$ is the evolution operator for the two-flavor problem with the modified matter potential 
$\tilde{V}$ and
\begin{equation}
\label{a8}
\beta = exp \left( -i \frac{1}{2} (\Delta_{31} + \Delta_{32})t - i \int_0^t dt' \left( V_{\mu e} (t') - \frac{3}{2} \tilde{V} (t') \right) \right) . 
\end{equation}
The quantity ${\bf U}_1$ is the solution of 
\begin{equation}
\label{a9}
i \frac{\partial}{\partial t} {\bf U}_1
= ({\bf U}^{\dagger}_0 {\cal H}_0 {\bf U}_0) {\bf U}_1 
\end{equation}
where 
\begin{equation}
\label{a10}
{\bf U}^{\dagger}_0 {\cal H}_0 {\bf U}_0 = 
\left(
\begin{array}{cc}
 0  &  J  \\
 J^{\dagger} &  0 
\end{array}
\right)
\end{equation}
with
\begin{equation}
\label{a11}
J =  \frac{1}{2}  \sin 2 \theta_{13}  V_{\mu e} {\bf S}^{\dagger} \left(
\begin{array}{c}
  1   \\
  0
\end{array}
\right) \beta . 
\end{equation}

A perturbative solution of Eq. (\ref{a9}) is 
\begin{equation}
\label{a12}
{\bf U}_1 (T) = \left(
\begin{array}{cc}
1 - \int_0^T dt \int_0^t dt' J(t)J^{\dagger} (t')  &   -i \int_0^T dt J(t)   \\
 -i \int_0^T dt J^{\dagger}(t)    &     1 - \int_0^T dt \int_0^t dt' J^{\dagger} (t)J (t') 
\end{array}
\right) + {\cal O} (\sin^3 2 \theta_{13}) .
\end{equation}
Consider the integral appearing in Eq. (\ref{a12}):
\begin{eqnarray}
\label{a13}
I(T) &=& \int_0^T dt J(t)  \\
&=& \frac{1}{2} \sin 2 \theta_{13} \int_0^T dt e^{-i (\Delta_{31} + \Delta_{32})t/2}  \exp 
\left(- i \int_0^t dt' \left( V_{\mu e} (t') - \frac{3}{2} \tilde{V} (t') \right) \right) V_{\mu e} {\bf S}^{\dagger} \left(
\begin{array}{c}
  1   \\
  0
\end{array}
\right)  . \nonumber
\end{eqnarray}
In this integral the first exponential $e^{-i (\Delta_{31} + \Delta_{32})t/2} $ is a rapidly oscillating function in the Sun, averaging to zero inside the integral. (Note that $\Delta_{32} R_{\odot} \sim 10^5$ for a 10 MeV neutrino). Indeed, Riemann-Lebesgue lemma ensures that 
\[
\int_a^b dt f(t) e^{ixg(t)} \rightarrow 0 
\]
as $x \rightarrow \infty$, provided that 
\[
\int_a^b dt |f(t)|
\]
exists. The leading contribution to the integral in Eq. (\ref{a13}) is \cite{bender}
\begin{equation}
\label{a14}
\int_a^b dt f(t) e^{ixg(t)} \sim \left. \frac{f(t)}{ix(dg/dt)} e^{ixg(t)} \right|_{t=a}^b 
\end{equation}
as $x\rightarrow \infty$. (For a proof of this statement with integration by parts see \cite{Balantekin:2003dc}). 
Using Eq. (\ref{a14}) to calculate the integral in Eq. (\ref{a13}) yields
\begin{equation}
\label{a15}
I (R_{\odot}) = -i  \sin 2 \theta_{13} \left( \frac{1}{\Delta_{32} + \Delta_{31}} V_{\mu e} (r=0) \right)
\left( \begin{array}{c}
  1   \\
  0
\end{array}
\right)  . 
\end{equation}
For a 10 MeV neutrino with $\delta m_{32}^2 \sim 2 \times 10^{-3}$ eV$^2$, assuming an electron density of 100 N$_{\rm A}$ / cm$^3$ in the neutrino production region, the quantity inside the parentheses in Eq. (\ref{a15}) 
\begin{equation}
\label{a16}
\alpha \equiv \frac{1}{\Delta_{32} + \Delta_{31}} V_{\mu e} (r=0) \sim 3 \times 10^{-2}. 
\end{equation}
If one takes $\alpha=0$, one finds, e.g. $\Psi_e$ on the surface of the Sun to be
\begin{equation}
\label{a17}
\Psi_e = \cos^2 \theta_{13} \psi_1 + \sin^2 \theta_{13} \beta, 
\end{equation}
where $\psi_1$ is the electron neutrino wave function obtained in two-flavor evolution with the modified matter potential 
$\tilde{V}$. Upon averaging over the Earth-Sun distance (see e.g.  \cite{Balantekin:1998yb}) 
one gets Eq. (\ref{b1}). Including the first order correction in ${\bf U}_1$, which is proportional t $\alpha$ of Eq. (\ref{a16}), gives 
\begin{equation}
\label{a18}
P_{3\times3}( \nu_e \rightarrow  \nu_e) = \cos^4{\theta_{13}}  ( 1 - 4 \sin^2\theta_{13} \alpha) \> 
P_{2\times2}( \nu_e \rightarrow  \nu_e \>{\rm with}\> N_e
\cos^2{\theta_{13}})  + \sin^4{\theta_{13}}  (1+ 4 \cos^2 \theta_{13} \alpha). 
\end{equation} 
A different derivation of this formula was given in Ref. \cite{Fogli:2001wi}. Even for the largest value of $\theta_{13}$ quoted by T2K experiment, this correction to the survival probability is less than  one percent for a 10 MeV neutrino. 

\section{CP-violating phase}

It is rather straightforward to show that the CP-violating phase factorizes out in the neutrino evolution Hamiltonian of Eq. (\ref{CProt}): 
\begin{equation}
\label{a19}
\tilde{\bf H} (\delta_{CP}) = {\bf S} \tilde{\bf H} (\delta_{CP}=0) {\bf S}^{\dagger}
\end{equation}
with
\begin{equation}
{\bf S} = \left(
\begin{array}{ccc}
 1 & 0  & 0  \\
 0 & 1  & 0  \\
 0 & 0  & e^{i \delta_{CP}}  
\end{array}
\right) .
\end{equation}
Eq. (\ref{a19}) implies that the neutrino evolution operator can be similarly factorized. This factorization gives us interesting sum rules:
 Electron neutrino survival probability, $P (\nu_e \rightarrow \nu_e)$ is independent of the value of the CP-violating phase, $\delta$; or equivalently, in the absence of sterile neutrino mixing, 
the combination $P (\nu_{\mu} \rightarrow \nu_e) + P (\nu_{\tau} \rightarrow \nu_e)$ at a fixed energy is independent of the value of the CP-violating phase \cite{Balantekin:2007es}. However the probability 
 $P (\nu_{\mu} \rightarrow \nu_e) $ depends on the $\delta_{CP}$. This is why the disappearance experiments such as those that use reactor antineutrinos can quote a probability independent of $\delta_{CP}$, whereas the appearance experiments, such as those that use accelerator neutrinos, cannot disentangle $\theta_{13}$ from $\delta_{CP}$. 
 It is possible to derive similar sum rules for other amplitudes \cite{Kneller:2009vd}. A discussion of the breakdown of this formula when sterile neutrinos are present is given in Ref. \cite{Palazzo:2011rj}. 

\section{Concluding Remarks} 

We provided a brief summary of the limits and preliminary measurements of the still poorly known neutrino mixing angle, $\theta_{13}$. Many analyses of the solar neutrino data utilize a formula connecting two- and three-flavor evolution. We gave a critical assessment of the validity of this formula. 

If the angle $\theta_{13}$ is rather large, there may be significant astrophysical implications. 
A brief review of the impact of the value of the mixing angle $\theta_{13}$ on various astrophysical phenomena was given in \cite{Balantekin:2010sv}.

\ack 
I thank A. Palazzo for discussions and correcting a typo in the earlier version of the manuscript.  
This work was supported in part
by the U.S. National Science Foundation Grant No. PHY-0855082
and
in part by the University of Wisconsin Research Committee with funds
granted by the Wisconsin Alumni Research Foundation.

\end{document}